\documentclass[aps,prb,reprint,superscriptaddress]{revtex4-2}
\usepackage{graphicx}
\usepackage{amsmath,amssymb}
\usepackage{bm}
\usepackage[colorlinks=true,linkcolor=blue,citecolor=blue,urlcolor=blue]{hyperref}

\begin{document}

\title{First-principles design of main-group dimer defects in ZnO as candidate quantum defects}

\author{Taejoon Park}
\affiliation{SKKU Advanced Institute of Nanotechnology, Sungkyunkwan University, Suwon, Gyeonggi 16419, Republic of Korea}
\affiliation{Department of Energy Systems Research and Department of Physics, Ajou University, Suwon 16499, Republic of Korea}
\affiliation{Department of Materials Science and Engineering, University of Wisconsin-Madison, Madison, Wisconsin 53706, USA}

\author{Erik Alfredo Perez}
\affiliation{Department of Materials Science and Engineering, University of Wisconsin-Madison, Madison, Wisconsin 53706, USA}

\author{Shimin Zhang}
\affiliation{Department of Materials Science and Engineering, University of Wisconsin-Madison, Madison, Wisconsin 53706, USA}

\author{Yuan Ping}
\email[Contact author: ]{yping3@wisc.edu}
\affiliation{Department of Materials Science and Engineering, University of Wisconsin-Madison, Madison, Wisconsin 53706, USA}
\affiliation{Department of Physics, University of Wisconsin-Madison, Madison, Wisconsin 53706, USA}
\affiliation{Department of Chemistry, University of Wisconsin-Madison, Madison, Wisconsin 53706, USA}

\author{Hosung Seo}
\email[Contact author: ]{seo.hosung@skku.edu}
\affiliation{SKKU Advanced Institute of Nanotechnology, Sungkyunkwan University, Suwon, Gyeonggi 16419, Republic of Korea}
\affiliation{Department of Energy Systems Research and Department of Physics, Ajou University, Suwon 16499, Republic of Korea}
\affiliation{Department of Materials Science and Engineering, University of Wisconsin-Madison, Madison, Wisconsin 53706, USA}
\affiliation{Department of Quantum Information Engineering, Sungkyunkwan University, Suwon, Gyeonggi 16419, Republic of Korea}

\date{\today}

\begin{abstract}
Zinc oxide (ZnO), a wide-band-gap semiconductor with mature growth techniques, is a promising host for optically active quantum spins. Yet, optically active quantum defects in ZnO remain largely unexplored. Here, we identify and characterize a family of double substitutional impurities in ZnO, formed by main-group donor-acceptor (DA) pairs, as candidates for optically active quantum defects. Using hybrid density functional theory (DFT), we systematically investigate double substitutional DA complexes and their defect physics, including electronic structure, thermodynamic stability, and optical properties. The proposed defects exhibit isolated defect states, strong spin localization on the acceptor site, and \emph{C\textsubscript{3v}} symmetry. Importantly, the electronic structure of the DA pairs is largely determined by the atomic properties of their constituent atoms. We further examine their optical characteristics, including zero-phonon lines (ZPLs), radiative lifetimes, and nonradiative decay to assess their viability as color centers. Notably, among the dimers, (Si\textsubscript{Zn}-B\textsubscript{O})\textsuperscript{+} and (Ge\textsubscript{Zn}-B\textsubscript{O})\textsuperscript{+} exhibit visible optical transitions with sub-microsecond radiative lifetimes and robust charge states against optical ionization, while (Si\textsubscript{Zn}-C\textsubscript{O})\textsuperscript{2+} shows the smallest Huang--Rhys factor, approximately 5.6. Our results propose a new family of main-group donor-acceptor defects in ZnO as promising candidates for optically active spin defects.
\end{abstract}

\maketitle

\section{Introduction}

Point defects in wide bandgap semiconductors have emerged as robust solid-state platforms for next-generation quantum technologies. Their localized electronic states can host electronic spins that are both optically addressable and coherently controllable. This dual capability underpins applications in quantum sensing, computing, and communication\cite{ref1,ref2,ref3}. The nitrogen-vacancy (NV) center in diamond\cite{ref4,ref5,ref6} and vacancy-related complexes in silicon carbide (SiC)\cite{ref7,ref8,ref9} have become benchmark defect systems, possessing spin-triplet ground states, well-defined optical transitions and long coherence time. These systems have already demonstrated essential quantum functionalities, including spin-photon entanglement\cite{ref10,ref11,ref12,ref13}, room-temperature quantum operation\cite{ref14,ref15,ref16}, spin qubit registers with quantum memory\cite{ref17,ref18} and nanoscale magnetometry\cite{ref19,ref20,ref21,ref22}.

Despite these achievements, existing platforms face several challenges, including the difficulty of large-scale growth of high-quality crystals\cite{ref23,ref24}, deterministic defect fabrication\cite{ref25} and qubit scalability\cite{ref26,ref27}. Identifying alternative quantum defects in technologically mature semiconductors could open new pathways toward scalable and integrable qubit architectures\cite{ref3,ref27,ref28,ref29}. The vast diversity of deep-level centers across semiconductors may offer a fertile ground for novel quantum functionalities such as tunable single-photon sources\cite{ref30,ref31,ref32,ref33} and quantum transduction\cite{ref34,ref35}. Exploring alternative quantum defects is also important from fundamental perspective as it may reveal new types of qubit operations\cite{ref36,ref37,ref38} and decoherence protection schemes\cite{ref39,ref40,ref41}. Together, these considerations highlight the need to expand the design space of quantum defects.

Among candidate host materials\cite{ref1,ref2,ref42,ref43,ref44,ref45,ref46,ref47}, zinc oxide (ZnO) has emerged as a promising oxide platform for optically active defect centers, with room-temperature single-photon emission and stable defect-related optical transitions\cite{ref28,ref48,ref49,ref50}. ZnO has a wide band gap of 3.4 eV, enabling intragap optical transitions in the visible range with minimal background absorption. It also contains a low natural abundance of spinful isotopes (\textsuperscript{67}Zn: 4.1\%, \textsuperscript{17}O: 0.04\%)\cite{ref51}, which suppresses nuclear-spin-induced decoherence even without isotopic purification. In addition, ZnO synthesis is highly mature, with a range of growth techniques compatible with industrial processes\cite{ref52}. Recent experiments have further demonstrated coherent spin control in ZnO, particularly in shallow donor-bound electrons created via gallium or indium implantation\cite{ref53,ref54,ref55}. In these systems, the longitudinal relaxation time (T\textsubscript{1}) exceeds 100 ms, while the Hahn-spin-echo coherence time (T\textsubscript{2}) reaches 50 \ensuremath{\mu}s\cite{ref53}, further supporting ZnO as a viable qubit platform.

Beyond shallow donors, identifying deep-level quantum defects in ZnO is especially important, as such defects can enable room-temperature operation along with quantum emission in the infrared and visible ranges\cite{ref49}. Notably, Zhang \emph{et al.} recently proposed a transition-metal (TM) vacancy complex (Mo--vacancy) that hosts deep-level spin states capable of high-fidelity single-shot readout at elevated temperatures\cite{ref56}. These defect candidates also exhibit bright visible emission with high quantum yield and an unusually small Huang--Rhys (HR) factor in ZnO, resulting in sharp zero-phonon lines. This work opens a promising avenue for high-performance deep-level qubits in ZnO and motivates further exploration of a broader class of defect candidates.

In this work, we expand the design space of deep-level defects in ZnO by investigating substitutional impurities composed of \emph{s}- and \emph{p}-block main-group elements using hybrid density functional theory (DFT) and spin-flip time-dependent DFT. We identify a new class of spin-qubit candidates consisting of donor--acceptor (DA) pairs forming dimer-like structures, with Al, Si, Ga, or Ge substituting at Zn sites and B or C substituting at O sites. Our results show that these dimers exhibit \ensuremath{C_{3v}} symmetry and spin-triplet ground states, with strongly localized spin density on the acceptor site. Detailed electronic-structure analysis reveals that the defect states are largely governed by the atomic orbitals of the constituent elements, providing a clear chemical design principle. The calculated defect formation energies are below 7 eV, suggesting potential experimental realizability. We further evaluate optical properties and find that (Si\textsubscript{Zn}-B\textsubscript{O})\textsuperscript{+} and (Ge\textsubscript{Zn}-B\textsubscript{O})\textsuperscript{+} exhibit sub-microsecond radiative lifetimes with suppressed nonradiative decay channels, indicating bright emission and high quantum yield. Notably, (Si\textsubscript{Zn}-C\textsubscript{O})\textsuperscript{2+} exhibits a relatively small Huang--Rhys factor (5.6), comparable to that of the Mo--vacancy center, while maintaining a high quantum yield. Overall, our results establish a new family of ZnO donor--acceptor dimer defects as promising candidates for optically active spin qubits.

\section{Results}

\subsection{Defect structure and electronic properties}
\begin{figure*}[t]
\includegraphics[width=0.95\textwidth]{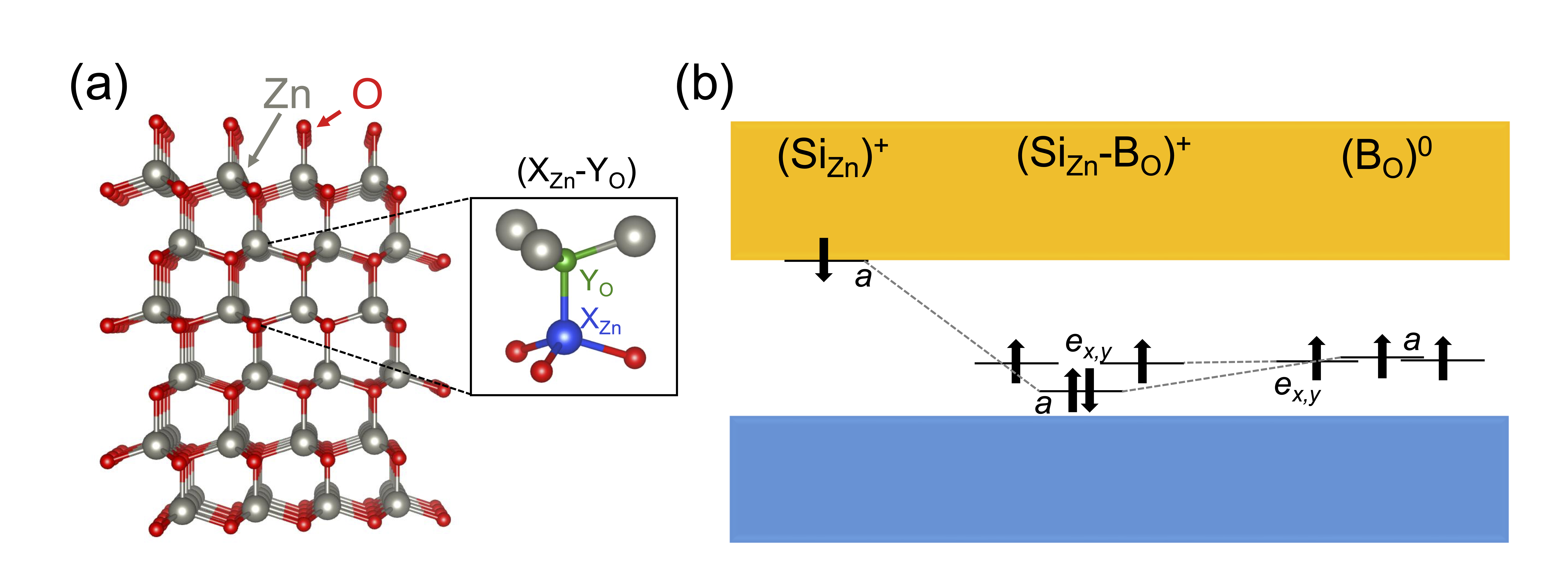}
\caption{Double substitutional defects in ZnO (a) Schematic illustration of dimer defects X\textsubscript{Zn}-Y\textsubscript{O}, where X = Al, Ga, Si, and Ge, and Y = B and C, encapsulated in ZnO. Gray and red spheres represent zinc and oxygen, respectively. Blue and green spheres indicate the impurity substituting Zn and O atoms, respectively. (b) Schematic of the defect level diagram of Si\textsubscript{Zn}B\textsubscript{O}\textsuperscript{+} dimer defect, illustrating the origin of the defect states. The blue and orange areas indicate the valence and conduction band of ZnO, respectively. The dotted gray lines represent the orbital contributions to each defect state.}
\label{fig1}
\end{figure*}

 Fig.~\ref{fig1}(a) illustrates the defect configurations investigated in this study. We focus on donor-acceptor (DA) type defect pairs\cite{ref57,ref58} with Ga, Al, Si, and Ge at the Zn site, while substitutional C and B are considered at the O site. Ga\textsubscript{Zn} and Al\textsubscript{Zn} are well-established shallow donors in ZnO\cite{ref53,ref54,ref55,ref59,ref60,ref61} while C\textsubscript{O} has been identified as a possible deep acceptor\cite{ref62}. To further expand the design space, we also consider Ge and Si as alternative donor species and B as an additional acceptor candidate for DA-type spin defects.

Fig.~\ref{fig1}(b) illustrates the microscopic origin of the defect orbitals for the (Si\textsubscript{Zn}-B\textsubscript{O})\textsuperscript{+} dimer. We first note that the isolated B\textsubscript{O} defect forms \emph{S} = 3/2 ground state in ZnO, with defect states located deep within the band gap. The \emph{p\textsubscript{x}} and \emph{p\textsubscript{y}} orbitals of the B atom primarily give rise to the doubly degenerate \emph{e\textsubscript{x}} and \emph{e\textsubscript{y}} states, which are strongly localized on the B site. In contrast, the \emph{a} state originates from hybridization between the Zn \emph{p\textsubscript{z}} orbital and a B \emph{sp}-like orbital. When a Si atom substitutes at a neighboring Zn site, the \emph{a} state is further lowered in energy due to hybridization between Si and B \emph{p} orbitals. This interaction results in a significant energy splitting between the \emph{a} and \emph{e} states in the dimer defect. We attribute this enhanced splitting to the relatively lower energy of the Si \emph{p} orbitals compared to those of Zn\cite{ref63}.

These results indicate that Si plays a dual role in the dimer defect: it acts as a donor while simultaneously lowering the energy of the \emph{a} defect level. As a result, the (Si\textsubscript{Zn}-B\textsubscript{O})\textsuperscript{+} defect hosts four electrons occupying the defect orbitals in an open-shell configuration, yielding a total spin S = 1 according to Hund's rule. The dimer preserves \emph{C\textsubscript{3v}} symmetry, and thus the degeneracy of the \emph{e} manifold is maintained. With this framework in place, we now proceed to analyze the detailed electronic structures of all dimer defects considered in this study.

\begin{figure*}[t]
\includegraphics[width=0.95\textwidth]{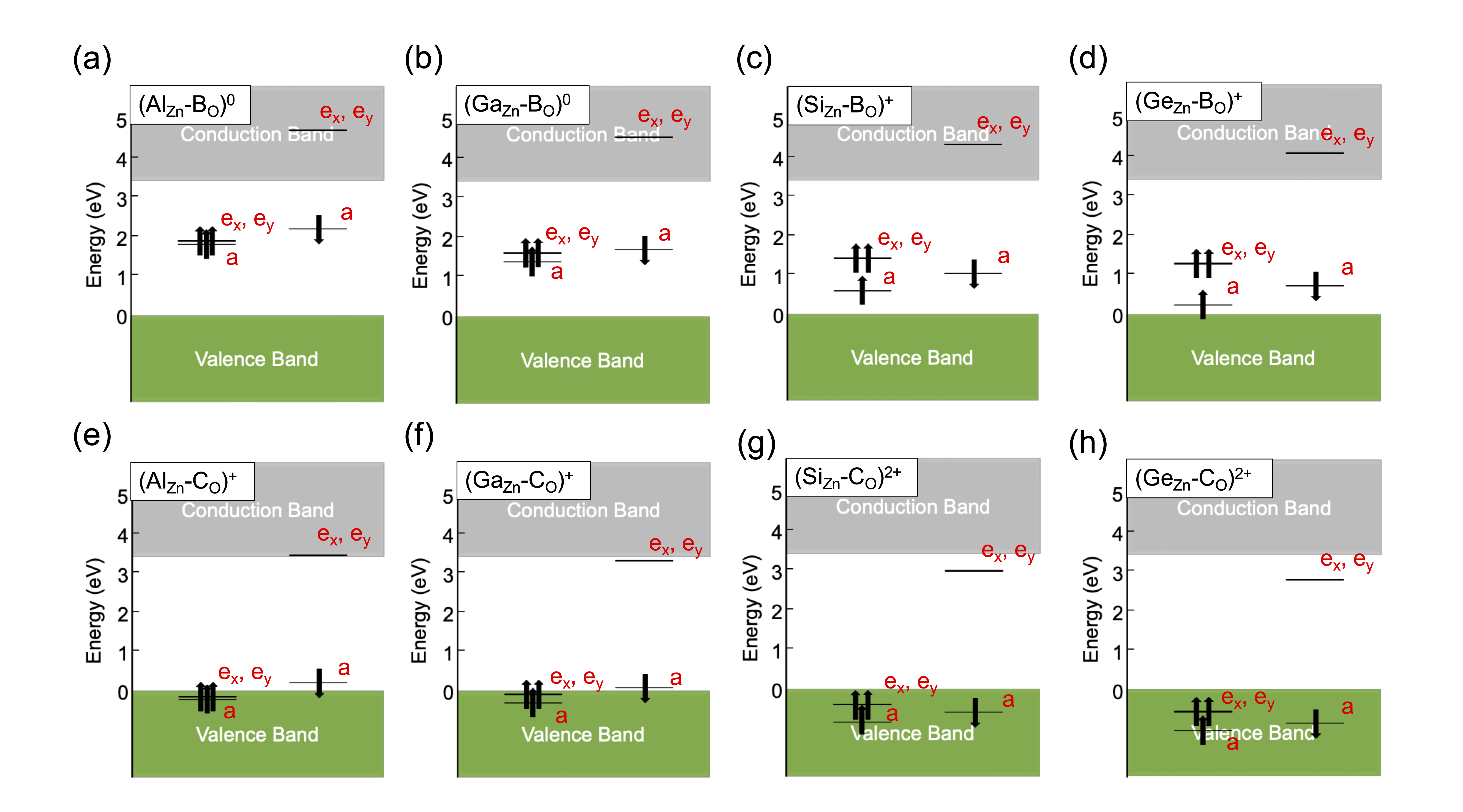}
\caption{Defect level diagrams of dimer defects in ZnO. (a-h) Computed defect level diagrams of spin-triplet boron-based dimer defects (a-d) and carbon-based defects (e-h). For (Al,Ga)\textsubscript{Zn}-B\textsubscript{O}, (Si,Ge)\textsubscript{Zn}-B\textsubscript{O}, (Al,Ga)\textsubscript{Zn}-C\textsubscript{O}, and (Si,Ge)\textsubscript{Zn}-B\textsubscript{O}, charge state of 0, +1, +1, and +2 are required to have spin-triplet ground states.}
\label{fig2}
\end{figure*}

Figs.~\ref{fig2}(a)--(d) present the spin-polarized defect-level diagrams of the boron-based dimers: (Al\textsubscript{Zn}-B\textsubscript{O})\textsuperscript{0}, (Ga\textsubscript{Zn}-B\textsubscript{O})\textsuperscript{0}, (Si\textsubscript{Zn}-B\textsubscript{O})\textsuperscript{+} and (Ge\textsubscript{Zn}-B\textsubscript{O})\textsuperscript{+}. All configurations exhibit spin-triplet (S = 1) ground states, characterized by a common set of \emph{a} and \emph{e} defect levels within the band gap, consistent with the orbital picture introduced in Fig.~\ref{fig1}(b). In the spin-majority channel, a non-degenerate \emph{a} level and two degenerate \emph{e\textsubscript{x}} and \emph{e\textsubscript{y}} levels reside within the ZnO band gap. In the spin-minority channel, the \emph{a} level remains in the band gap, while the degenerate \emph{e} levels lie above the conduction band minimum (CBM). The C\textsubscript{3v} symmetry is preserved across all configurations, maintaining the degeneracy of the \emph{e} manifold. The spin density is consistently localized on the boron atom (Supplementary Fig. 1a--1d), confirming the acceptor-dominated character of the defect states.

A systematic evolution of the electronic structure is observed as the donor atom is varied. Replacing Al with Ga, Si, and Ge shifts the \emph{a} defect levels toward the valence band maximum (VBM) by 0.43 eV, 1.13 eV, and 1.36 eV, respectively. A similar but smaller downward shift is observed for the spin-minority \emph{e} levels, amounting to 0.19 eV, 0.37 eV, and 0.62 eV, respectively. The \emph{a} level exhibits a stronger sensitivity to donor substitution than the \emph{e} levels, due to its significant contribution from the donor \emph{p}\textsubscript{z} orbital. In contrast, the \emph{e} levels are primarily derived from the B \emph{p}\textsubscript{x} and \emph{p}\textsubscript{y} orbitals and are therefore less sensitive to the donor species. The observed trends can be understood in terms of atomic orbital energetics. Substituting Al with Si increases the nuclear charge, resulting in a substantial lowering of the defect orbital energies. In the case of Ga, the 4\emph{p} orbitals lie lower in energy than the Al 3\emph{p} orbitals due to the poor shielding effect of the filled 3\emph{d} shell, which leads to a higher effective nuclear charge\cite{ref63}. A similar downward shift is observed when Si is replaced by Ge, reflecting the increased nuclear charge within the same group. Overall, the variation in defect level positions follows the relative energies of the constituent atomic orbitals.

Figs.~\ref{fig2}(e)--(h) present the calculated defect-level diagrams for dimers with carbon substituting at the anion site (C-based dimers). We consider specific charge states that yield the same electronic configuration as the (Si\textsubscript{Zn}-B\textsubscript{O})\textsuperscript{+} defect: +1 for Al\textsubscript{Zn}-C\textsubscript{O} and Ga\textsubscript{Zn}-C\textsubscript{O}, and +2 for Si\textsubscript{Zn}-C\textsubscript{O} and Ge\textsubscript{Zn}-C\textsubscript{O}. Under these conditions, all C-based dimers exhibit spin-triplet (S = 1) ground states. These defects also retain C\textsubscript{3v} symmetry, with spin density predominantly localized on the carbon site, similar to the behavior observed in the boron-based dimers (Supplementary Fig. 1e--1h).

Despite these similarities, the C-based dimers show notable differences in their defect-level structures compared to the B-based systems. In particular, the occupied \emph{a} and \emph{e} states in the spin-majority channel are located below the valence band maximum (VBM) for all C-based dimers. In the spin-minority channel, the unoccupied degenerate \emph{e} states lie close to the conduction band minimum (CBM). For (Al\textsubscript{Zn}-C\textsubscript{O})\textsuperscript{+} (Fig.~\ref{fig2}(e)), the \emph{e} levels are positioned approximately 15 meV above the CBM. These levels shift progressively downward for (Ga\textsubscript{Zn}-C\textsubscript{O})\textsuperscript{+}, (Si\textsubscript{Zn}-C\textsubscript{O})\textsuperscript{2+}, and (Ge\textsubscript{Zn}-C\textsubscript{O})\textsuperscript{2+} (Fig.~\ref{fig2}(f)--(h)), lying 0.19 eV, 0.40 eV, and 0.54 eV below the CBM, respectively. The overall downward shift of both \emph{a} and \emph{e} levels relative to the B-based dimers can be attributed to the larger nuclear charge of carbon compared to boron. As in the B-based systems, replacing the donor atom from Al to Ga, Si, or Ge leads to a further lowering of the \emph{a} level, consistent with the trends discussed above. These observations indicate that the defect-level positions in C-based dimers follow similar trends with donor substitution, while exhibiting a systematic offset due to the different acceptor species.

\subsection{Defect energetics}
 To evaluate the stability of the dimer complexes in ZnO, we computed their defect formation energy as a function of the Fermi level \(\epsilon_{F}\) within the ZnO band gap according to the standard formalism\cite{ref64}:
\begin{multline}
E^{f}\left\lbrack X^{q} \right\rbrack = E_{tot}\left\lbrack X^{q} \right\rbrack - E_{tot}\lbrack \mathrm{pristine}\rbrack \\ - \sum_{i}{n_{i}\mu_{i}} + q(E_{VBM} + \epsilon_{F}) + E_{corr}
\end{multline}
where \(E_{tot}\left\lbrack X^{q} \right\rbrack\) and \(E_{tot}\lbrack pristine\rbrack\) are the total energies of the defect-containing and pristine supercells, respectively. \(n_{i}\) and \(\mu_{i}\) denote the number and chemical potentials of species \emph{i}, referenced to their stable phases (B\textsubscript{2}O\textsubscript{3}, C, Al\textsubscript{2}O\textsubscript{3}, SiO\textsubscript{2}, Ga\textsubscript{2}O\textsubscript{3} and Ge). \(E_{VBM}\) is the VBM and \(E_{corr}\) accounts for finite-size corrections in charged supercells\cite{ref65}.

\begin{figure}[b]
\includegraphics[width=\columnwidth]{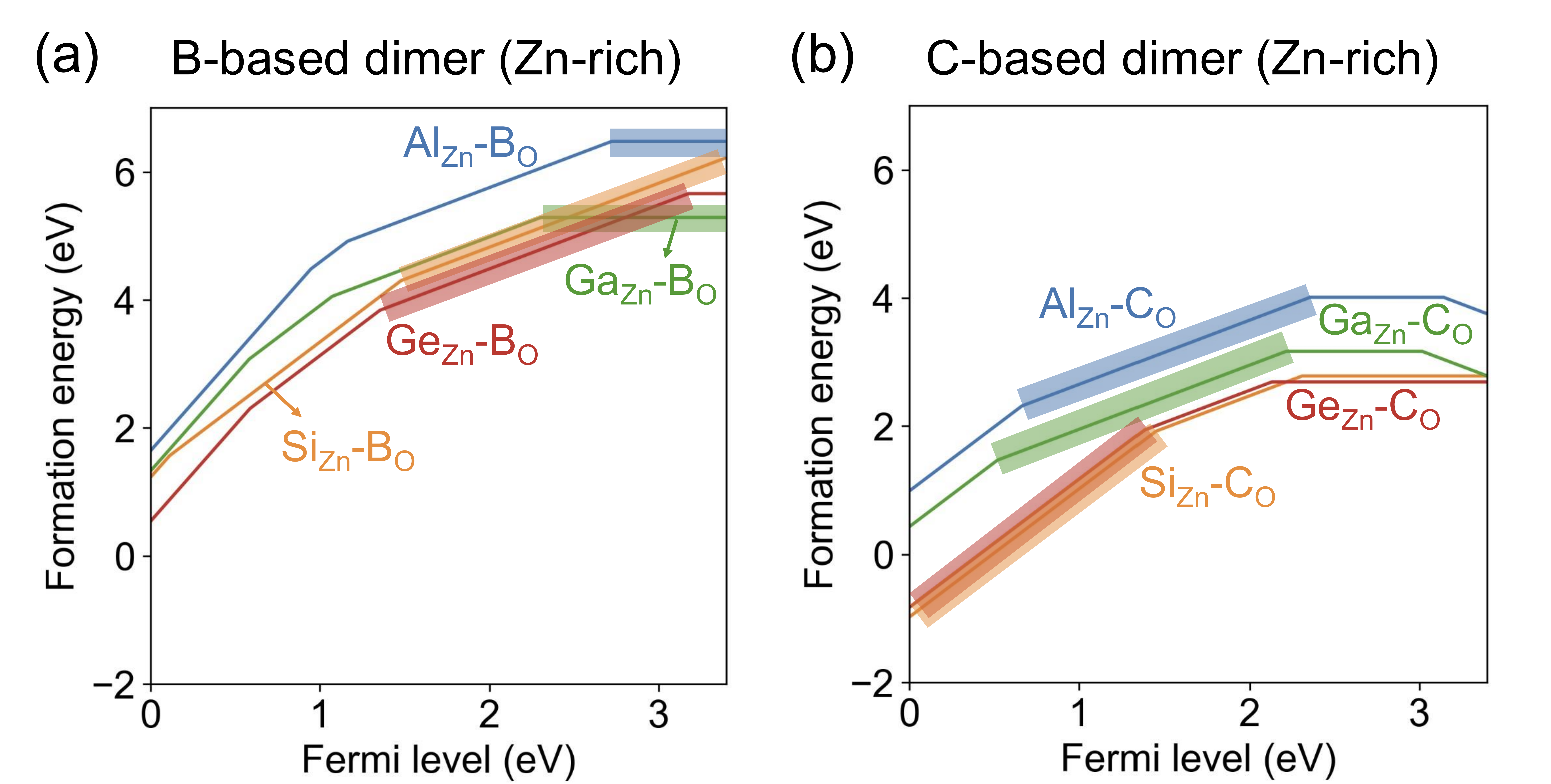}
\caption{Energetics of the dimer defects in ZnO. (a,b) Computed defect formation energy of B-based (a) and C-based (b) dimer defects under the Zn-rich condition. Defect formation energies under the O-rich condition are provided in Supplementary Fig. 2. The color bar indicates the stability region for the dimer defects with the spin-triplet state.}
\label{fig3}
\end{figure}

Figs.~\ref{fig3}(a) and (b) present the formation energies of the B-based and C-based dimers, respectively, under Zn-rich condition. The color bars indicate the Fermi-level ranges over which the spin-triplet charge states are stable. For B-based dimers, these regions lie in the upper part of the band gap, corresponding to \emph{n}-type conditions. In contrast, C-based dimers stabilize their spin-triplet states in mid-gap to lower Fermi-level regions, corresponding to intrinsic to \emph{p}-type conditions. While ZnO is typically \emph{n}-type, achieving \emph{p}-type conductivity remains challenging, although notable progress has been reported\cite{ref66}. These results suggest that realization of the spin-triplet dimers may require control over the Fermi level, particularly for the C-based systems.

The calculated formation energies of all dimer defects are below 7 eV under Zn-rich conditions. For B-based dimers, (Al\textsubscript{Zn}-B\textsubscript{O})\textsuperscript{0} and (Ga\textsubscript{Zn}-B\textsubscript{O})\textsuperscript{0} exhibit formation energies of 6.48 eV and 5.29 eV, respectively, within their relevant Fermi-level ranges. (Si\textsubscript{Zn}-B\textsubscript{O})\textsuperscript{+} and (Ge\textsubscript{Zn}-B\textsubscript{O})\textsuperscript{+} show formation energies that vary with \(\epsilon_{F}\), ranging from 3--5 eV and 4--6 eV, respectively. For the C-based dimers, (Al\textsubscript{Zn}-C\textsubscript{O})\textsuperscript{+} and (Ga\textsubscript{Zn}-C\textsubscript{O})\textsuperscript{+} remain below 4 eV across the relevant Fermi-level range. Notably, (Si\textsubscript{Zn}-C\textsubscript{O})\textsuperscript{2+} and (Ge\textsubscript{Zn}-C\textsubscript{O})\textsuperscript{2+} exhibit formation energies below 2 eV, indicating relatively strong thermodynamic stability. For context, native point defects in ZnO typically have formation energies on the order of several eV\textsuperscript{67, 68}, while the NV center in diamond and vacancy-related defects in SiC have reported formation energies in the ranges of 4--6 eV and 7--8 eV, respectively\cite{ref1,ref69,ref70}. In this context, the formation energies of the proposed dimer defects fall within a similar range, suggesting that their experimental realization may be feasible.

\begin{figure*}[t]
\includegraphics[width=0.95\textwidth]{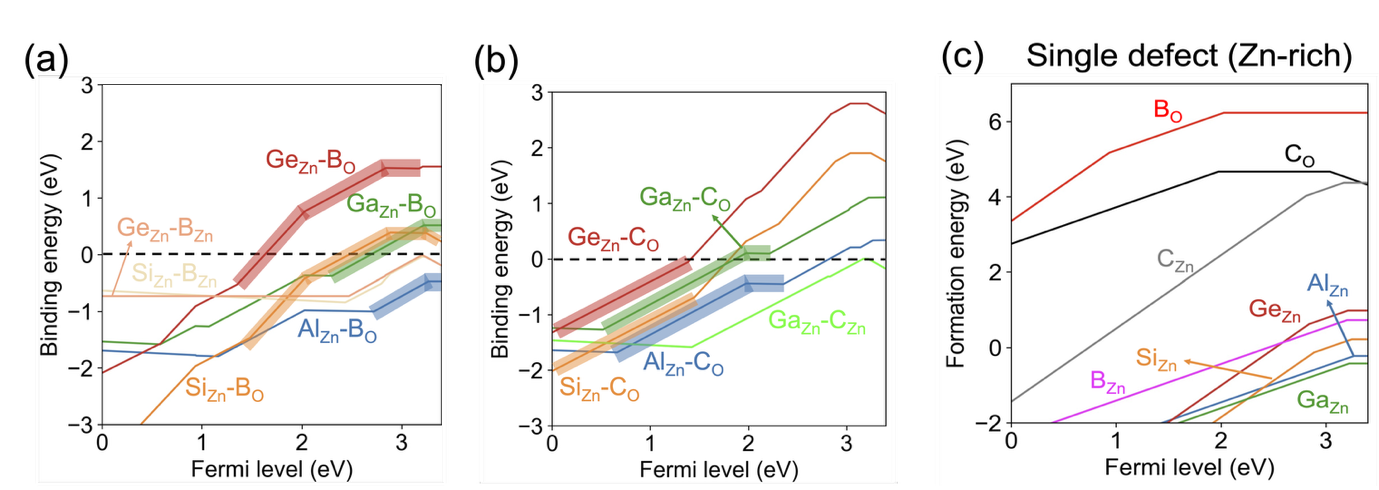}
\caption{Binding energy of dimer defects in ZnO. (a,b) Binding energy of B-based (a) and C-based (b) dimers as a function of the Fermi level. The horizontal dashed line separates positive and negative binding energy regions. Color bars indicate the Fermi-level range where the dimer is in the spin-triplet (S = 1) state. (c) Defect formation energies of single-site substitutional defects.}
\label{fig4}
\end{figure*}

By using the calculated defect formation energies, we further evaluate the binding energies of the dimer defects in ZnO, as shown in Figs.~\ref{fig4}(a) and (b). The binding energy is defined as \cite{ref64}:
\begin{equation}
E_{bind} = \ E^{f}\lbrack A\rbrack + E^{f}\lbrack B\rbrack - E^{f}\lbrack AB\rbrack
\end{equation}
where \(E^{f}\lbrack AB\rbrack\) is the defect formation energy of the DA complex and \(E^{f}\lbrack A\rbrack,\ \ E^{f}\lbrack B\rbrack\) are the defect formation energies of the individual single substitutional defects. The computed formation energies of single-site substitutional defects are shown in Fig. 4c, including Zn-site substitution (Al\textsubscript{Zn}, Ga\textsubscript{Zn}, Si\textsubscript{Zn}, Ge\textsubscript{Zn}, B\textsubscript{Zn}, C\textsubscript{Zn}) and O-site substitution (B\textsubscript{O} and C\textsubscript{O}). All Zn-site substitutions exhibit relatively low formation energies and act as donor-like defects. Notably, for B and C, substitution at the Zn site is energetically more favorable than at the O site. These trends are consistent with previous experimental and theoretical studies on donor-type defects in ZnO.\cite{ref53,ref55,ref59,ref60,ref71}.

Fig.~\ref{fig4}(a) shows the binding energies of B-based dimers as a function of the Fermi level. The (Al\textsubscript{Zn}--B\textsubscript{O}) defect exhibits negative binding energy across the entire Fermi-level range, indicating that formation of this complex from isolated defects is energetically unfavorable. In contrast, (Si\textsubscript{Zn}--B\textsubscript{O}), (Ga\textsubscript{Zn}--B\textsubscript{O}), and (Ge\textsubscript{Zn}--B\textsubscript{O}) exhibit positive binding energies over a broad Fermi-level range that overlaps with the stability region of their spin-triplet states.

Fig.~\ref{fig4}(b) presents the binding energies of the C-based dimers. For (Al\textsubscript{Zn}--C\textsubscript{O}), (Si\textsubscript{Zn}--C\textsubscript{O}), and (Ge\textsubscript{Zn}--C\textsubscript{O}), the binding energies are negative in the lower Fermi-level region (\emph{p}-type conditions), where the spin-triplet states are stable. However, the binding energies become positive as the Fermi level approaches the CBM. This indicates that these complexes are more likely to form under \emph{n}-type conditions, although achieving the charge states required for spin-triplet configurations would then require additional control of the Fermi level.

We also examine alternative configurations in which both impurities occupy Zn sites, such as (Si\textsubscript{Zn}-B\textsubscript{Zn}), (Ge\textsubscript{Zn}-C\textsubscript{Zn}), and (Ga\textsubscript{Zn}-C\textsubscript{Zn}). These configurations exhibit lower formation energies than their corresponding Zn--O dimers (see Supplementary Fig. 3). However, their binding energies remain negative over a wide Fermi-level range, indicating that the interaction between the constituent defects is effectively repulsive. Overall, while O-site acceptor defects such as B\textsubscript{O} and C\textsubscript{O} have relatively higher formation energies compared to their Zn-site counterparts, they can form stable complexes with donor defects under \emph{n}-type conditions due to favorable binding energies.

\subsection{Optical properties of dimer defects in ZnO}
 Table~\ref{tab1} summarizes the computed vertical excitation energies (E\textsubscript{V}) and ZPLs for the \emph{a}\ensuremath{\rightarrow}\emph{e} defect-level transition of (Si\textsubscript{Zn}-B\textsubscript{O})\textsuperscript{+}, (Ge\textsubscript{Zn}-B\textsubscript{O})\textsuperscript{+}, (Si\textsubscript{Zn}-C\textsubscript{O})\textsuperscript{2+}, and (Ge\textsubscript{Zn}-C\textsubscript{O})\textsuperscript{2+}, obtained from constrained-DFT (cDFT) calculations\cite{ref72}. For Al- and Ga-based dimers, the cDFT calculations do not converge due to band inversion during the constrained calculation procedure. For the B-based dimers, the computed ZPL values of (Si\textsubscript{Zn}-B\textsubscript{O})\textsuperscript{+} and (Ge\textsubscript{Zn}-B\textsubscript{O})\textsuperscript{+} are 1.84 eV and 1.94 eV, respectively, while those for (Si\textsubscript{Zn}-C\textsubscript{O})\textsuperscript{2+} and (Ge\textsubscript{Zn}-C\textsubscript{O})\textsuperscript{2+} are 2.04 eV and 2.00 eV, respectively. All the ZPL energies lie within the visible spectral range.

\begin{table}[b]
\caption{\label{tab1}%
Computed optical properties of dimer defects in ZnO. Zero-phonon line (ZPL), Stokes-shift energy E\textsubscript{S} and vertical excitation energy E\textsubscript{V} are computed by cDFT. The ZPL is evaluated as E\textsubscript{V}-E\textsubscript{S}. Squared transition dipole moment \ensuremath{\mu}\textsuperscript{2} and radiative lifetime \ensuremath{\tau}\textsubscript{R} are computed within the RPA using single particle wavefunctions computed at HSE (\ensuremath{\alpha}=0.375).}
\begin{ruledtabular}
\begin{tabular}{lccccc}
Defect & ZPL (eV) & $E_S$ (eV) & $E_V$ (eV) & $\mu^2$ (bohr$^2$) & $\tau_R$ ($\mu$s) \\
\hline
(Si$_{\rm Zn}$-B$_{\rm O}$)$^{+}$ & 1.84 & 0.25 & 2.07 & 0.60 & 0.09 \\
(Ge$_{\rm Zn}$-B$_{\rm O}$)$^{+}$ & 1.94 & 0.24 & 2.18 & 0.55 & 0.07 \\
(Si$_{\rm Zn}$-C$_{\rm O}$)$^{2+}$ & 2.04 & 0.17 & 2.21 & -- & -- \\
(Ge$_{\rm Zn}$-C$_{\rm O}$)$^{2+}$ & 2.00 & 0.17 & 2.17 & -- & -- \\
\end{tabular}
\end{ruledtabular}
\end{table}

We also compute the polarization-resolved optical absorption spectra within the random-phase approximation (RPA), using single-particle wavefunctions obtained from HSE (\ensuremath{\alpha}=0.375) calculations (see Supplementary Fig. 4a, 4b). This analysis is important because an optically addressable defect should exhibit an optical transition that is distinguishable from bulk band-edge absorption. For the (Si\textsubscript{Zn}-B\textsubscript{O})\textsuperscript{+} and (Ge\textsubscript{Zn}-B\textsubscript{O})\textsuperscript{+} defects, the lowest RPA absorption peaks appear at 1.91 and 2.17 eV, respectively, which are close to the cDFT vertical excitation energies as shown in Table~\ref{tab1}, and show in-plane (x,y)-polarized character. This is consistent with the \emph{a}\ensuremath{\rightarrow}\emph{e} defect-level transition expected from \emph{C\textsubscript{3v}} symmetry.

In contrast, for the C-based dimers, no absorption peaks are resolved near the cDFT excitation energies; the dominant RPA features correspond to bulk-to-bulk transitions near the band edges (Supplementary Fig. 4c, 4d). This difference can be traced to the defect-level structure shown in Fig.~\ref{fig2}. The RPA spectrum is constructed from the ground-state single-particle states, in which the occupied \emph{a} and \emph{e} levels of the C-based dimers lie below the VBM. The \emph{a}\ensuremath{\rightarrow}\emph{e} excitation considered in the cDFT calculation therefore has no counterpart among the single-particle transitions, and the only transitions below the band gap is the VBM\ensuremath{\rightarrow}\emph{e} transition, which carries negligible oscillator strength. On the other hand, in the cDFT calculation, the excitation energy is obtained as a total-energy difference and includes the orbital relaxation of the defect states. For (Si\textsubscript{Zn}-C\textsubscript{O})\textsuperscript{2+} and (Ge\textsubscript{Zn}-C\textsubscript{O})\textsuperscript{2+}, the cDFT vertical excitation energies are 2.21 and 2.17 eV (Table~\ref{tab1}). The lowest RPA transitions arise from the VBM to the minority-spin \emph{e} level and occur at approximately 3.0 and 2.9 eV (see Fig.~\ref{fig2}). The cDFT energies thus lie 0.7--0.8 eV below any transition in the RPA spectrum. A correspondence between the cDFT and RPA excitation energies therefore cannot be established for the C-based dimers, and we exclude them from the quantitative comparison.

We estimate the brightness of the defect emission by computing its radiative lifetime (\ensuremath{\tau}\textsubscript{R}), for which we employed the Fermi's Golden rule under the dipole approximation\cite{ref73,ref74}.
\begin{equation}
\tau_{R} = \frac{3\pi\epsilon_{0}\hslash^{4}c^{3}}{n_{D}e^{2}E^{3}\mu^{2}}
\end{equation}
where \emph{c} is the speed of light, E is the excitation energy, \(\mu^{2}\) is the modulus square of transition dipole moment and \(n_{D}\) is reflective index computed from the dielectric constant of pristine ZnO. Computed radiative lifetime of (Si\textsubscript{Zn}-B\textsubscript{O})\textsuperscript{+} and (Ge\textsubscript{Zn}-B\textsubscript{O})\textsuperscript{+} are in the sub-microsecond regime, with values of 0.09 \ensuremath{\mu}s and 0.07 \ensuremath{\mu}s, respectively, as summarized in Table~\ref{tab1}. These lifetimes are longer than that of the NV center in diamond (\textasciitilde12 ns)\cite{ref75}, but significantly shorter than the boron vacancy center in h-BN (\textasciitilde20 \ensuremath{\mu}s)\cite{ref76}. The squared dipole moments for (Si\textsubscript{Zn}-B\textsubscript{O})\textsuperscript{+} and (Ge\textsubscript{Zn}-B\textsubscript{O})\textsuperscript{+} are 0.60 and 0.55 respectively. These values are smaller than that of NV center in diamond (\ensuremath{\mu}\textsuperscript{2} \textasciitilde{} 3.87)\cite{ref77}, but substantially larger than those of boron vacancy center in h-BN (\ensuremath{\mu}\textsuperscript{2} \textasciitilde{} 10\textsuperscript{-4})\cite{ref78} and the Mo-vacancy in ZnO (\ensuremath{\mu}\textsuperscript{2} \textasciitilde{} 10\textsuperscript{-2})\cite{ref56}.

In addition, to assess whether the defect can support a possible intersystem-crossing (ISC) pathway, we investigate the excited state energetics of a representative defect (Ge\textsubscript{Zn}-B\textsubscript{O})\textsuperscript{+}. We employ time-dependent DFT with the hybrid functional to compute the E\textsubscript{V} from the \textsuperscript{3}A\textsubscript{2} ground state to the triplet excited state \textsuperscript{3}E and the singlet states \textsuperscript{1}E and \textsuperscript{1}A\textsubscript{1}, as implemented in the WEST code\cite{ref79}. Spin-conserving TDDFT is used to obtain the triplet excited state, while spin-flip TDDFT is employed to access the singlet states. The computed \textsuperscript{1}E, \textsuperscript{1}A\textsubscript{1}, and \textsuperscript{3}E states lie 0.66 eV, 1.38 eV, and 2.05 eV above the \textsuperscript{3}A\textsubscript{2} ground state, respectively. This level structure is consistent with that of the NV center\cite{ref6,ref80,ref81} and suggests the presence of an ISC pathways mediated by intermediate singlet shelving states.

\subsection{Huang-Rhys factor and quantum yield of dimer defects}
\begin{figure}[t]
\includegraphics[width=\columnwidth]{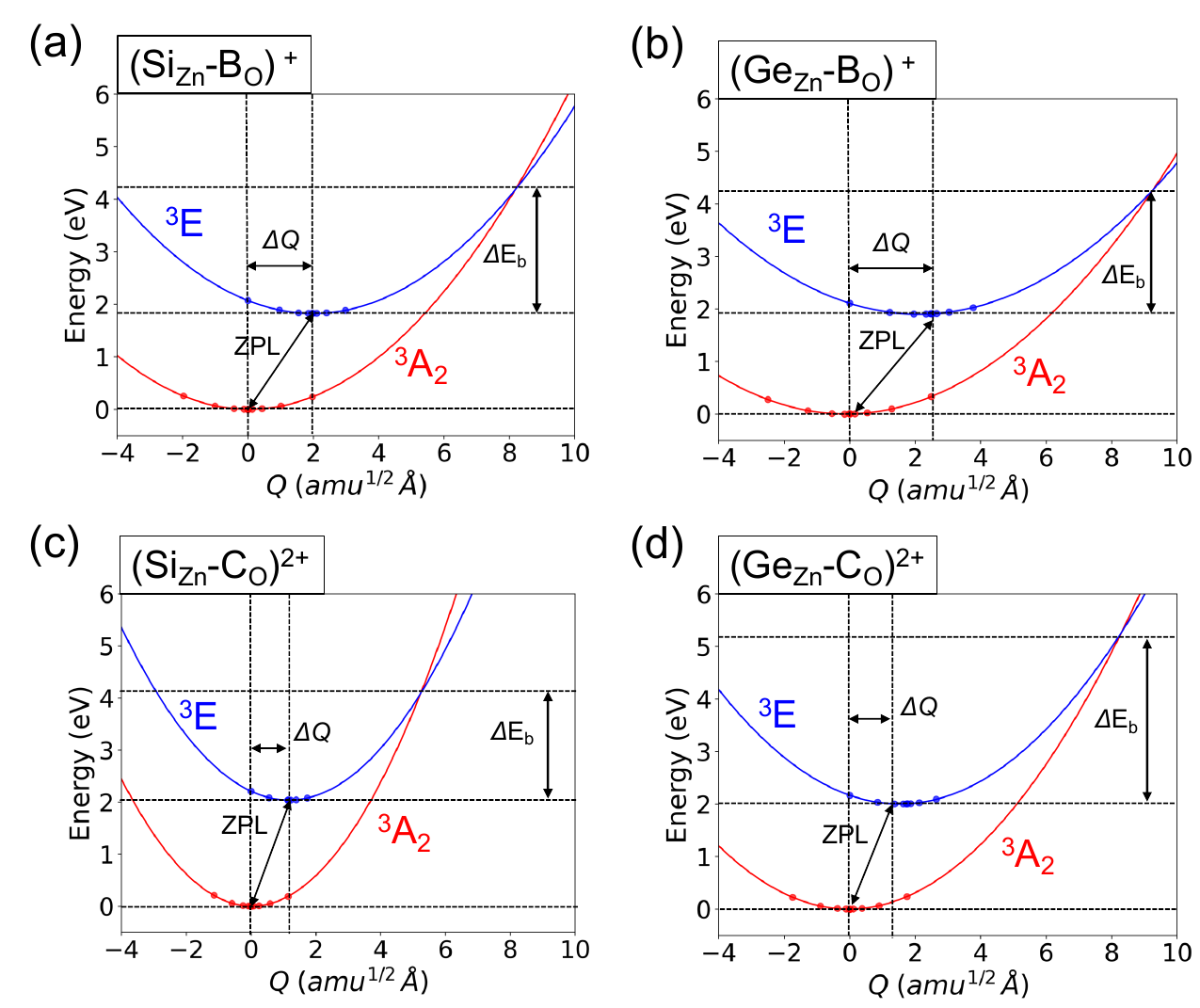}
\caption{Configuration coordinate diagram of the dimer structures in ZnO. (a-d) Potential energy surface of (Si\textsubscript{Zn}-B\textsubscript{O})\textsuperscript{+} (a), (Ge\textsubscript{Zn}-B\textsubscript{O})\textsuperscript{+} (b) (Si\textsubscript{Zn}-C\textsubscript{O})\textsuperscript{2+} (c) and (Ge\textsubscript{Zn}-C\textsubscript{O})\textsuperscript{2+} (d) in ZnO. PES is fitted by harmonic potentials with effective phonon modes. For all the dimer defects, classical barriers for the nonradiative process (\ensuremath{\Delta}E\textsubscript{b}) are larger than 2 eV which indicates the low possibility of the nonradiative decay in these defect systems.}
\label{fig5}
\end{figure}

 To characterize phonon-related optical properties of the dimer defects, we compute Huang-Rhys (HR) factor (S)\cite{ref82,ref83,ref84}, Debye-Waller factor (DW)\cite{ref85,ref86} nonradiative lifetime (\(\tau_{nr}\))\cite{ref87} and quantum yield (QY = \(\frac{1/\tau_{r}}{1/\tau_{r} + 1/\tau_{nr}}\)) of (Si\textsubscript{Zn}-B\textsubscript{O})\textsuperscript{+}, (Ge\textsubscript{Zn}-B\textsubscript{O})\textsuperscript{+}, (Si\textsubscript{Zn}-C\textsubscript{O})\textsuperscript{2+} and (Ge\textsubscript{Zn}-C\textsubscript{O})\textsuperscript{2+}. HR factor quantifies the strength of the electron-phonon coupling and the lattice relaxation associated with the electronic transition. DW factor estimates the relative intensity of the ZPL compared to the phonon sideband. HR factor and DW factor are computed as follows:
\begin{equation}
S = \omega_{eff}{\Delta Q}^{2}/2\hslash
\end{equation}\begin{equation}
DW = e^{- S}
\end{equation}
where \(\Delta Q\) and \(\omega_{eff}\) are mass-weighted displacements and effective phonon mode frequency, respectively. For the computation of \(\Delta Q\) and \(\omega_{eff}\), we plot the configuration coordinate diagram (CCD) of the defect system, which is shown in Fig.~\ref{fig5}. Within CCD, potential energy surface (PES) of ground and excited states is fit to harmonic potentials, allowing estimation of the effective phonon energy \ensuremath{\hbar}\ensuremath{\omega}\textsubscript{eff}.

In the 1D configuration coordinate model, \ensuremath{\Delta}Q between the equilibrium configurations of the two states is defined as follows\cite{ref6,ref72}:
\begin{equation}
\Delta Q = \ \sqrt{\sum_{\alpha,i}^{}{m_{\alpha}{\Delta R}_{\alpha,i}^{2}}}
\end{equation}
where \(m_{\alpha}\) is the atomic mass of atom \(\alpha\) and \({\Delta R}_{\alpha,i}^{\ }\) is the displacement between excited state and ground state (\(i = \{ x,\ y,\ z\})\).

To measure how fast the nonradiative recombination happens, we compute the nonradiative lifetime (\(\tau_{nr}\)), which is defined as follows under the static coupling approximation with one-dimensional (1D) phonon approximation\cite{ref74,ref86}:
\begin{equation}
\frac{1}{\tau_{nr}} = \frac{2\pi}{\hslash}g\left| W_{if} \right|^{2}X_{if}(T)
\end{equation}\begin{equation}
W_{if} = < \psi_{i}\left( \mathbf{r,R} \right)\left| \frac{\partial H}{\partial Q} \right|\psi_{f}\left( \mathbf{r,R} \right) > |_{R = R_{a}}
\end{equation}\begin{multline}
X_{if} = \sum_{n,m}{p_{in}{\left| \left\langle \phi_{fm}\left( \mathbf{R} \right)\left| Q - Q_{a} \right|\phi_{in}\left( \mathbf{R} \right) \right\rangle \right|}^{2}} \\
\times\ \delta(m\hslash\omega_{f} - n\hslash\omega_{i} + \Delta E_{if})
\end{multline}
Eq. (7) is separated into the electronic term (\(W_{if}\) in Eq. 8) and the phonon term (\(X_{if}\) in Eq. 9). \(g\) is the degeneracy factor of the final state and it depends on the number of equivalent atomic configuration. \(W_{if}\) term depends on the electronic wave function (\emph{\ensuremath{\psi}}) overlap and it is determined using finite difference of Kohn-Sham orbitals. \(X_{if}\) term describes the strength of the phonon contribution, and this term includes the energy conservation between the initial and final vibronic states, characterized by vibrational frequencies \(\omega_{i}\) and \(\omega_{f}\). Here, \(\phi\) denotes the phonon wavefunction, which is derived from harmonic oscillator. \(p_{in}\) is the occupation number of the vibronic state \(|\phi_{in} >\) following the Boltzmann distribution. Detailed derivations are provided in Refs.~\cite{ref86,ref87}.

Table~\ref{tab2} summarizes the phonon-related optical properties of the four dimer defects. All four defects exhibit high quantum yields close to unity, indicating strongly suppressed nonradiative decay. Interestingly, however, their \(\Delta Q\), HR factor, and DW factors differ substantially, suggesting that their photoluminescence spectra should have distinct phonon-sideband contributions.

\begin{table*}[t]
\caption{\label{tab2}%
Huang-Rhys factor and quantum yield of dimer defects in ZnO. \(\Delta Q\) indicates the mass-weighted displacements between the ground- and excited-state geometries. \emph{\ensuremath{\hbar}\ensuremath{\omega}\textsubscript{eff}} is the effective phonon energy. \emph{k} represents the number of phonons needed to mediate the transition. \emph{S} and DW denote Huang-Rhys and Debye-Waller factors, respectively. \(\tau_{NR}\) is the nonradiative lifetime, with \(W_{if}\) and \(X_{if}\), representing the electronic and phonon contribution to \(\tau_{NR}\). QY denotes the quantum yield.}
\begin{ruledtabular}
\begin{tabular}{lccccccccc}
Defect & $\Delta Q$ & $\hbar\omega_{\rm eff}$ & $k$ & $S$ & DW & $\tau_{NR}$ & $W_{if}$ & $X_{if}$ & QY \\
 & (amu$^{1/2}$\AA) & (meV) & & & & ($\mu$s) & (eV/amu$^{1/2}$\AA) & (amu\,\AA$^2$/eV) & \\
\hline
(Si$_{\rm Zn}$-B$_{\rm O}$)$^{+}$ & 1.97 & 22.95 & 80 & 10.69 & $2.28\times10^{-5}$ & Not allowed & $3.29\times10^{-3}$ & $4.17\times10^{-24}$ & High \\
(Ge$_{\rm Zn}$-B$_{\rm O}$)$^{+}$ & 2.49 & 20.13 & 97 & 14.98 & $3.12\times10^{-7}$ & Not allowed & $1.58\times10^{-3}$ & $1.76\times10^{-24}$ & High \\
(Si$_{\rm Zn}$-C$_{\rm O}$)$^{2+}$ & 1.15 & 35.48 & 58 & 5.57 & $3.83\times10^{-3}$ & Not allowed & $3.46\times10^{-3}$ & $2.42\times10^{-20}$ & High \\
(Ge$_{\rm Zn}$-C$_{\rm O}$)$^{2+}$ & 1.75 & 25.20 & 80 & 9.20 & $1.04\times10^{-4}$ & Not allowed & $2.48\times10^{-3}$ & $3.60\times10^{-26}$ & High \\
\end{tabular}
\end{ruledtabular}
\end{table*}

We find that the C-based dimers exhibit substantially weaker vibronic coupling than their B-based counterparts. The HR factor \emph{S} decreases from 10.69 to 5.57 for the Si-related dimers and from 14.98 to 9.20 for the Ge-related dimers upon replacing B with C, consistent with reduced structural reorganization during the optical transition. This reduction is accompanied by larger DW factors, indicating a greater relative contribution of the ZPL to the emission. In both B- and C-based dimers, the Ge-related complex exhibits stronger vibronic coupling than the Si-related counterpart, reflecting the larger mass-weighted displacement associated with the heavier Ge atom. Notably, (Si\textsubscript{Zn}-C\textsubscript{O})\textsuperscript{2+} combines the two favorable trends and shows the smallest mass-weighted displacement and HR factor (\emph{S} = 5.57) among the four dimers, suggesting weaker phonon sidebands and potentially better-resolved ZPL emission.

We note that nonradiative decay is strongly suppressed for all four dimers despite their substantial lattice relaxation. This suppression can be understood from the large number of phonons required to dissipate the electronic transition energy\cite{ref88}, with \emph{k} = 58--97 substantially exceeding the corresponding HR factors, S = 5.57--14.98. This large difference indicates that the initial and final vibrational wavefunctions overlap only weakly at the energy required for the transition, consistent with the small vibrational factors \emph{X\textsubscript{if}}\hspace{0pt} obtained for all four dimers.

The configuration coordinate diagram (CCD) in Fig.~\ref{fig5} provides a complementary physical picture of the suppressed nonradiative decay obtained in Table~\ref{tab2}. Within the semiclassical configuration coordinate picture, the crossing between the ground state PES (red curve) and the excited state PES (blue curve) defines an activation barrier, \ensuremath{\Delta}E\textsubscript{b}, measured from the minimum of the excited state PES to the crossing region. The calculated \ensuremath{\Delta}E\textsubscript{b} values are 2.35, 2.37, 2.75, and 3.27 eV for (Si\textsubscript{Zn}--B\textsubscript{O})\textsuperscript{+}, (Ge\textsubscript{Zn}--B\textsubscript{O})\textsuperscript{+}, (Si\textsubscript{Zn}--C\textsubscript{O})\textsuperscript{2+}, and (Ge\textsubscript{Zn}--C\textsubscript{O})\textsuperscript{2+}, respectively. Such large barriers make thermal access to the crossing region unfavorable, and therefore provide a semiclassical indication of suppressed nonradiative relaxation. This semiclassical picture is consistent with the small \emph{X\textsubscript{if}} values in Table~\ref{tab2}, which reflect the vibrational contribution to the nonradiative transition. Taken together, these results clarify why substantial lattice relaxation can coexist with strongly suppressed nonradiative decay. Although the sizable HR factors imply pronounced phonon sidebands, the small vibrational overlap factors \emph{X\textsubscript{if}} (Table~\ref{tab2}) and the large crossing barriers (Fig.~\ref{fig5}) together ensure that radiative decay remains the dominant relaxation channel, resulting in near-unity quantum yields.

\subsection{Ionization energies of the dimer defects}
 Quantum spin defects are often initialized and read out by optical pumping processes. We assess the robustness of the defect charge and spin states under optical pumping by evaluating the ionization energy (IE) and electron capture energy (CE). The IE is the energy required to remove an electron from the defect, while the CE corresponds to excitation of an electron from the valence band maximum (VBM) to the defect state. For stable intra-defect optical cycling, both CE and IE should exceed the ZPL energy.

\begin{table}[b]
\caption{\label{tab3}%
Electron capture energy and ionization energy of the dimer defects in ZnO. Computed electron capture energy and ionization energy. Hyphen-minus (-) sign denotes that the corresponding CE or IE are larger than the band gap of ZnO.}
\begin{ruledtabular}
\begin{tabular}{cccc}
Defect & CE / IE (eV) & Defect & CE / IE (eV) \\
\hline
(Al$_{\rm Zn}$-B$_{\rm O}$)$^{0}$ & -- / 0.68 & (Si$_{\rm Zn}$-B$_{\rm O}$)$^{+}$ & -- / 1.91 \\
(Ga$_{\rm Zn}$-B$_{\rm O}$)$^{0}$ & -- / 1.09 & (Ge$_{\rm Zn}$-B$_{\rm O}$)$^{+}$ & 3.18 / 2.05 \\
(Al$_{\rm Zn}$-C$_{\rm O}$)$^{+}$ & 2.36 / 2.74 & (Si$_{\rm Zn}$-C$_{\rm O}$)$^{2+}$ & 1.45 / -- \\
(Ga$_{\rm Zn}$-C$_{\rm O}$)$^{+}$ & 2.22 / 2.88 & (Ge$_{\rm Zn}$-C$_{\rm O}$)$^{2+}$ & 1.38 / -- \\
\end{tabular}
\end{ruledtabular}
\end{table}

We computed the CE and IE from the energy differences between the relevant charge transition levels (CTLs) and the band edges, as summarized in Table~\ref{tab3}. For each defect, CE and IE are defined with respect to the charge state hosting the spin-triplet ground state. In Al\textsubscript{Zn}-B\textsubscript{O}, for example, whose spin-triplet ground state occurs in the neutral charge state, CE is defined as the energy difference between the (-1/0) CTL and the VBM, and IE as that between the (0/+1) CTL and the CBM.

Table~\ref{tab3} reveals a general trend: defects whose spin-triplet ground state occurs in a more positive charge state exhibit larger IEs but smaller CEs. Ionization becomes more difficult because an electron must be removed from an already positively charged defect, whereas electron capture becomes more accessible because the defect can lower its charge state by accepting an electron. Charge-state robustness should therefore be judged from CE and IE together, rather than from either quantity alone.

For the B-based dimers, the cation-site atom primarily controls the IE. Al\textsubscript{Zn}-B\textsubscript{O} and Ga\textsubscript{Zn}-B\textsubscript{O}, whose spin-triplet ground states occur in the neutral charge state, show small IEs of 0.68 and 1.09 eV, while their CEs exceed the ZnO band gap. For Si\textsubscript{Zn}-B\textsubscript{O} and Ge\textsubscript{Zn}-B\textsubscript{O}, whose spin-triplet ground state is the +1 charge state, the IE increases to 1.91 and 2.05 eV, consistent with their occupied defect levels lying deeper below the CBM (Fig.~\ref{fig2}). Their CEs remain large because the high-lying (0/+1) CTLs make electron capture into the neutral charge state energetically costly. Thus, Si\textsubscript{Zn}-B\textsubscript{O} and Ge\textsubscript{Zn}-B\textsubscript{O} resist ionization more strongly while retaining large electron-capture energies. The C-based dimers follow the same trend at more positive charge states. Al\textsubscript{Zn}-C\textsubscript{O} and Ga\textsubscript{Zn}-C\textsubscript{O}, with +1 spin-triplet ground states, exhibit balanced CE and IE values of 2.36/2.74 and 2.22/2.88 eV, respectively. For Si\textsubscript{Zn}-C\textsubscript{O} and Ge\textsubscript{Zn}-C\textsubscript{O}, with +2 spin-triplet ground states, ionization to the +3 charge state is energetically costly, giving IEs larger than the ZnO band gap, whereas the (+1/+2) CTLs lie closer to the VBM, reducing the CEs to 1.45 and 1.38 eV.

Overall, the proposed dimers exhibit distinct charge-state stability profiles under optical pumping: the neutral B-based dimers are mainly limited by relatively small IEs, whereas the +2 C-based dimers are mainly limited by relatively small CEs. The most robust candidates are therefore (Si\textsubscript{Zn}-B\textsubscript{O})\textsuperscript{+}, (Ge\textsubscript{Zn}-B\textsubscript{O})\textsuperscript{+}, (Al\textsubscript{Zn}-C\textsubscript{O})\textsuperscript{+}, and (Ga\textsubscript{Zn}-C\textsubscript{O})\textsuperscript{+}, which retain sufficiently large CE and IE to preserve the spin-triplet charge state under optical excitation.

\section{Conclusion}

We propose a class of optically addressable defects in ZnO based on donor--acceptor (DA) complexes formed by double-substitutional main-group impurities. Using first-principles hybrid density functional theory, we systematically investigate their electronic structures, thermodynamic stability, and optical properties. The proposed complexes exhibit spin-triplet ground states, strong spin localization on the acceptor site, and \emph{C\textsubscript{3v}} symmetry. Across the dimer family, the defect-level positions and orbital characters are systematically governed by the atomic \emph{s}- and \emph{p}-orbital energetics of the constituent impurities.

The optical calculations reveal distinct characteristics across the dimer family. (Si\textsubscript{Zn}-B\textsubscript{O})\textsuperscript{+} and (Ge\textsubscript{Zn}-B\textsubscript{O})\textsuperscript{+} exhibit visible defect--defect transitions together with favorable optical charge-state stability, while their relatively large HR factors indicate strong vibronic coupling. By contrast, (Si\textsubscript{Zn}-C\textsubscript{O})\textsuperscript{2+} has the smallest mass-weighted displacement and HR factor among the dimers (S = 5.57), indicating reduced vibronic coupling within this family. Importantly, the magnitude of lattice relaxation does not directly determine the efficiency of nonradiative recombination. The nonradiative decay is strongly suppressed, allowing radiative emission to remain dominant with high quantum yields.

Notably, the excited state analysis of (Ga\textsubscript{Zn}-C\textsubscript{O})\textsuperscript{+} is complicated by defect-level reordering, which prevents convergence of the target excited state within conventional cDFT. We therefore estimate its optical transition energies using a Hubbard-U-assisted cDFT approach (Supplementary Note 1), obtaining approximate ZPL and vertical excitation energies of 1.5 and 1.7 eV, respectively.

The thermodynamic calculations also identify specific challenges for experimental realization. For (Si\textsubscript{Zn}-C\textsubscript{O})\textsuperscript{2+}, dimer association is unfavorable over the Fermi-level range where the desired 2+ charge state is stable, although it becomes favorable under n-type conditions. This suggests that the complex should first be formed under favorable conditions for association and subsequently brought into the desired charge state. On the other hand, the B-based dimers face a different limitation. B\textsubscript{O} is substantially less favorable to form than B\textsubscript{Zn}. Therefore, accessing the required O site configuration may require nonequilibrium growth. Explicit calculations of migration, site exchange, association, and dissociation barriers will be needed to assess these possibilities.

Overall, our results establish systematic relationships between the constituent main-group impurities and the electronic, thermodynamic, and optical properties of DA complexes in ZnO. Further characterization of spin-dependent excited-state processes, including spin--orbit coupling and intersystem crossing, together with defect-formation kinetics, will be important for assessing their experimental realization and optical control.

\section{Methods}

\subsection{First-principles calculations}
We performed DFT calculations to simulate the geometric and electronic structure of the defect immersed in the lattice of the ZnO host. We used the Vienna Ab initio Simulation Package (VASP)\cite{ref89} and employed a plane-wave basis set with an energy cutoff of 400 eV. We applied the screened hybrid functional of Heyd-Scuseria-Ernzerhof (HSE)\cite{ref90} together with the projector-augmented wave (PAW) method\cite{ref91} as implemented in VASP. The mixing parameter was set to 0.375, which successfully reproduces the band gap of ZnO. The range-separation parameter was fixed at the standard HSE06 value of 0.2 \AA{}\textsuperscript{-1}.

To simulate isolated defect centers, we used a 192-atom supercell with Gamma-only k-point sampling. The convergence criterion for self-consistent calculations was set to 10\textsuperscript{-5} eV. Atomic positions were relaxed until the residual forces became smaller than 0.01 eV/\AA{}. The spin density was calculated and visualized using VASPKIT\cite{ref92}. In the computation of defect formation energies, we employed the Freysoldt-Neugebauer-Van de Walle (FNV) correction scheme\cite{ref65} for charge correction, using the experimental dielectric constant epsilon = 8.1\cite{ref93}.

We employed constrained DFT as implemented in VASP to calculate excitation energies between triplet states by fixing the occupations of the defect levels. Partial occupations were used in the cDFT calculations to avoid occupation-dependent symmetry breaking\cite{ref94,ref95}. For the Hubbard \emph{U}-assisted cDFT calculations, we used the Dudarev scheme\cite{ref96} as implemented in VASP.

The optical absorption spectrum was calculated within the random-phase approximation (RPA) using the Yambo code\cite{ref97,ref98} including local-field effects in the polarizability. Single-particle states were generated using the plane-wave DFT code Quantum ESPRESSO\cite{ref99} with ONCV pseudopotentials\cite{ref100,ref101} and a wavefunction cutoff of 80 Ry.

Vertical excitation energies of the triplet and singlet excited states are computed using spin-conserving and spin-flip TDDFT, as implemented in the WEST code\cite{ref79,ref102,ref103}. Supercell-size convergence tests for TDDFT calculations are provided in the Supplementary Fig. 8.

We used the NONRAD code for the computation of nonradiative recombination rates\cite{ref104}.

\section*{Data availability}
The data that support the findings of this article, including the first-principles
input and output files and the post-processing scripts, are available from the
corresponding authors upon reasonable request.

\begin{acknowledgments}
This study is supported by the National Research Foundation (NRF) of Korea grant funded by the Korean government (MSIT) (No. 2023R1A2C1006270 and No. RS-2025-25454922), by Creation of the Quantum Information Science R\&D Ecosystem (Grant No. RS-2023-NR068116) through the NRF of Korea funded by the Korea government (MSIT). This work was supported by Institute of Information \& communications Technology Planning \& Evaluation (IITP) grant funded by the Korea government (MSIT) (RS-2025-25464252). This work was supported by the National Supercomputing Center with supercomputing resources including technical support (KSC-2024-CRE-0526). This research was supported by the education and training program of the Quantum Information Research Support Center, funded through the National research foundation of Korea (NRF) by the Ministry of science and ICT (MSIT) of the Korean government (No.2021M3H3A103657313). Ping, Perez, and Zhang acknowledge the support by the National Science Foundation under grant no. DMR-2143233 and the support by AFOSR CFIRE program under Grant No. FA9550-23-1-0418. During the preparation of this work the authors used ChatGPT (GPT-5.6, OpenAI) and Claude (Fable 5, Anthropic) in order to assist with code debugging and verification and to improve the readability of the manuscript. After using these tools, the authors reviewed and edited the content as needed and take full responsibility for the content of the publication.
\end{acknowledgments}


\begin{thebibliography}{104}
\bibitem{ref1} J. R. Weber, W. F. Koehl, J. B. Varley, A. Janotti, B. B. Buckley, C. G. Van de Walle, and D. D. Awschalom, Quantum computing with defects. Proc. Natl. Acad. Sci. U.S.A. 107, 8513-8518 (2010).

\bibitem{ref2} G. Wolfowicz, F. J. Heremans, C. P. Anderson, S. Kanai, H. Seo, A. Gali, G. Galli, and D. D. Awschalom, Quantum guidelines for solid-state spin defects. Nat. Rev. Mater. 6, 906-925 (2021).

\bibitem{ref3} D. D. Awschalom, R. Hanson, J. Wrachtrup, and B. B. Zhou, Quantum technologies with optically interfaced solid-state spins. Nat. Photonics 12, 516-527 (2018).

\bibitem{ref4} F. Jelezko, T. Gaebel, I. Popa, A. Gruber, and J. Wrachtrup, Observation of coherent oscillations in a single electron spin. Phys. Rev. Lett\emph{.} 92, 076401 (2004).

\bibitem{ref5} J. Wrachtrup and F. Jelezko, Processing quantum information in diamond. J. Phys.: Condens. Matter 18, S807-S824 (2006).

\bibitem{ref6} M. W. Doherty, N. B. Manson, P. Delaney, F. Jelezko, J. Wrachtrup, and L. C. L. Hollenberg, The nitrogen-vacancy colour centre in diamond. Phys. Rep. 528, 1-45 (2013).

\bibitem{ref7} J. R. Weber, W. F. Koehl, J. B. Varley, A. Janotti, B. B. Buckley, C. G. Van de Walle, and D. D. Awschalom, Defects in SiC for quantum computing. J. Appl. Phys. 109, 102417 (2011).

\bibitem{ref8} D. J. Christle, A. L. Falk, P. Andrich, P. V. Klimov, J. Ul Hassan, N. T. Son, E. Janzén, T. Ohshima, and D. D. Awschalom, Isolated electron spins in silicon carbide with millisecond coherence times. Nat. Mater. 14, 160-163 (2015).

\bibitem{ref9} N. T. Son, C. P. Anderson, A. Bourassa, K. C. Miao, C. Babin, M. Widmann, M. Niethammer, J. Ul Hassan, N. Morioka, I. G. Ivanov, F. Kaiser, J. Wrachtrup, and D. D. Awschalom, Developing silicon carbide for quantum spintronics. Appl. Phys. Lett. 116, 190501 (2020).

\bibitem{ref10} M. Pompili, S. L. N. Hermans, S. Baier, H. K. C. Beukers, P. C. Humphreys, R. N. Schouten, R. F. L. Vermeulen, M. J. Tiggelman, L. dos Santos Martins, B. Dirkse, S. Wehner, and R. Hanson, Realization of a multi-node quantum network of remote solid-state qubits. Science 372, 259-264 (2021).

\bibitem{ref11} C. T. Nguyen, D. D. Sukachev, M. K. Bhaskar, B. Machielse, D. S. Levonian, E. N. Knall, P. Stroganov, R. Riedinger, H. Park, M. Lončar, and M. D. Lukin, Quantum network nodes based on diamond qubits with an efficient nanophotonic interface. Phys. Rev. Lett. 123, 183602 (2019).

\bibitem{ref12} E. Togan, Y. Chu, A. S. Trifonov, L. Jiang, J. Maze, L. Childress, M. V. G. Dutt, A. S. Sørensen, P. R. Hemmer, A. S. Zibrov, and M. D. Lukin, Quantum entanglement between an optical photon and a solid-state spin qubit. Nature 466, 730-734 (2010).

\bibitem{ref13} B. Hensen, H. Bernien, A. E. Dréau, A. Reiserer, N. Kalb, M. S. Blok, J. Ruitenberg, R. F. L. Vermeulen, R. N. Schouten, C. Abellán, W. Amaya, V. Pruneri, M. W. Mitchell, M. Markham, D. J. Twitchen, D. Elkouss, S. Wehner, T. H. Taminiau, and R. Hanson, Loophole-free Bell inequality violation using electron spins separated by 1.3 kilometres. Nature 526, 682-686 (2015).

\bibitem{ref14} E. D. Herbschleb, H. Kato, Y. Maruyama, T. Danjo, T. Makino, S. Yamasaki, I. Ohki, K. Hayashi, H. Morishita, M. Fujiwara, and N. Mizuochi, Ultra-long coherence times amongst room-temperature solid-state spins. Nat. Commun. 10, 3766 (2019).

\bibitem{ref15} W. F. Koehl, B. B. Buckley, F. J. Heremans, G. Calusine, and D. D. Awschalom, Room temperature coherent control of defect spin qubits in silicon carbide. Nature 479, 84-87 (2011).

\bibitem{ref16} J.-F. Wang, F.-F. Yan, Q. Li, Z.-H. Liu, H. Liu, G.-P. Guo, L.-P. Guo, X. Zhou, J.-M. Cui, J. Wang, Z.-Q. Zhou, X.-Y. Xu, J.-S. Xu, C.-F. Li, and G.-C. Guo, Coherent control of nitrogen-vacancy center spins in silicon carbide at room temperature. Phys. Rev. Lett. 124, 223601 (2020).

\bibitem{ref17} C. E. Bradley, J. Randall, M. H. Abobeih, R. C. Berrevoets, M. J. Degen, M. A. Bakker, M. Markham, D. J. Twitchen, and T. H. Taminiau, A ten-qubit solid-state spin register with quantum memory up to one minute. Phys. Rev. X 9, 031045 (2019).

\bibitem{ref18} F. J. González and R. Coto, Decoherence-protected quantum register of nuclear spins in diamond. Quantum Sci. Technol. 7, 025015 (2022).

\bibitem{ref19} L. Rondin, J.-P. Tetienne, T. Hingant, J.-F. Roch, P. Maletinsky, and V. Jacques, Magnetometry with nitrogen-vacancy defects in diamond. Rep. Prog. Phys. 77, 056503 (2014).

\bibitem{ref20} B. J. Maertz, A. P. Wijnheijmer, G. D. Fuchs, M. E. Nowakowski, and D. D. Awschalom, Vector magnetic field microscopy using nitrogen vacancy centers in diamond. Appl. Phys. Lett. 96, 092504 (2010).

\bibitem{ref21} L. M. Pham, D. Le Sage, P. L. Stanwix, T. K. Yeung, D. Glenn, A. Trifonov, P. Cappellaro, P. R. Hemmer, M. D. Lukin, H. Park, A. Yacoby, and R. L. Walsworth, Magnetic field imaging with nitrogen-vacancy ensembles. New J. Phys. 13, 045021 (2011).

\bibitem{ref22} A. Wickenbrock, H. Zheng, L. Bougas, N. Leefer, S. Afach, A. Jarmola, V. M. Acosta, and D. Budker, Microwave-free magnetometry with nitrogen-vacancy centers in diamond. Appl. Phys. Lett. 109, 053505 (2016).

\bibitem{ref23} L. V. H. Rodgers, L. B. Hughes, M. Xie, P. C. Maurer, S. Kolkowitz, A. C. Bleszynski Jayich, and N. P. de Leon, Materials challenges for quantum technologies based on color centers in diamond. MRS Bull. 46, 623-633 (2021).

\bibitem{ref24} C. Becher, W. Gao, S. Kar, C. D. Marciniak, T. Monz, J. G. Bartholomew, P. Goldner, H. Loh, E. Marcellina, K. E. J. Goh, T. S. Koh, B. Weber, Z. Mu, J.-Y. Tsai, Q. Yan, T. Huber-Loyola, S. Höfling, S. Gyger, S. Steinhauer, and V. Zwiller, 2023 roadmap for materials for quantum technologies. Mater. Quantum Technol. 3, 012501 (2023).

\bibitem{ref25} D. Lee and J. A. Gupta, Perspectives on deterministic control of quantum point defects by scanned probes. Nanophotonics 8, 2033-2040 (2019).

\bibitem{ref26} M. Atatüre, D. Englund, N. Vamivakas, S.-Y. Lee, and J. Wrachtrup, Material platforms for spin-based photonic quantum technologies. Nat. Rev. Mater. 3, 38-51 (2018).

\bibitem{ref27} C. P. Anderson and D. D. Awschalom, Embracing imperfection for quantum technologies. Phys. Today 76, 26-33 (2023).

\bibitem{ref28} I. Aharonovich, D. Englund, and M. Toth, Solid-state single-photon emitters. Nat. Photonics 10, 631-641 (2016).

\bibitem{ref29} E. Altman, K. R. Brown, G. Carleo, L. D. Carr, E. Demler, C. Chin, B. DeMarco, S. E. Economou, M. A. Eriksson, K.-M. C. Fu, M. Greiner, K. R. A. Hazzard, R. G. Hulet, A. J. Kollár, B. L. Lev, M. D. Lukin, R. Ma, X. Mi, S. Misra, C. Monroe, K. Murch, Z. Nazario, K.-K. Ni, A. C. Potter, P. Roushan, M. Saffman, M. Schleier-Smith, I. Siddiqi, R. Simmonds, M. Singh, I. B. Spielman, K. Temme, D. S. Weiss, J. Vučković, V. Vuletić, J. Ye, and M. Zwierlein, Quantum simulators: Architectures and opportunities. PRX Quantum 2, 017003 (2021).

\bibitem{ref30} T. T. Tran, K. Bray, M. J. Ford, M. Toth, and I. Aharonovich, Quantum emission from hexagonal boron nitride monolayers. Nat. Nanotechnol. 11, 37-41 (2016).

\bibitem{ref31} G. Grosso, H. Moon, B. Lienhard, S. Ali, D. K. Efetov, M. M. Furchi, P. Jarillo-Herrero, M. J. Ford, I. Aharonovich, and D. Englund, Tunable and high-purity room-temperature single-photon emission from atomic defects in hexagonal boron nitride. Nat. Commun. 8, 705 (2017).

\bibitem{ref32} A. M. Berhane, K.-Y. Jeong, Z. Bodrog, S. Fiedler, T. Schröder, N. Vico Triviño, T. Palacios, A. Gali, M. Toth, D. Englund, and I. Aharonovich, Bright room-temperature single-photon emission from defects in gallium nitride. Adv. Mater. 29, 1605092 (2017).

\bibitem{ref33} Y. Zhou, Z. Wang, A. Rasmita, S. Kim, A. Berhane, Z. Bodrog, G. Adamo, A. Gali, I. Aharonovich, and W.-B. Gao, Room-temperature solid-state quantum emitters in the telecom range. Sci. Adv. 4, eaar3580 (2018).

\bibitem{ref34} C. O'Brien, N. Lauk, S. Blum, G. Morigi, and M. Fleischhauer, Interfacing superconducting qubits and telecom photons via a rare-earth-doped crystal. Phys. Rev. Lett. 113, 063603 (2014).

\bibitem{ref35} E. Janitz, M. K. Bhaskar, and L. Childress, Cavity quantum electrodynamics with color centers in diamond. Optica 7, 1232-1252 (2020).

\bibitem{ref36} X. Gao, S. Vaidya, K. Li, Z. Ge, S. Dikshit, S. Zhang, P. Ju, K. Shen, Y. Jin, Y. Ping, and T. Li, Single nuclear spin detection and control in a van der Waals material. Nature 643, 943-949 (2025).

\bibitem{ref37} N.-J. Guo, S. Li, W. Liu, Y.-Z. Yang, X.-D. Zeng, S. Yu, Y. Meng, Z.-P. Li, Z.-A. Wang, L.-K. Xie, R.-C. Ge, J.-F. Wang, Q. Li, J.-S. Xu, Y.-T. Wang, J.-S. Tang, A. Gali, C.-F. Li, and G.-C. Guo, Coherent control of an ultrabright single spin in hexagonal boron nitride at room temperature. Nat. Commun. 14, 2893 (2023).

\bibitem{ref38} S. C. Scholten, P. Singh, A. J. Healey, I. O. Robertson, G. Haim, C. Tan, D. A. Broadway, L. Wang, H. Abe, T. Ohshima, M. Kianinia, P. Reineck, I. Aharonovich, and J.-P. Tetienne, Multi-species optically addressable spin defects in a van der Waals material. Nat. Commun. 15, 6727 (2024).

\bibitem{ref39} J. Lee, H. Park, and H. Seo, First-principles theory of extending the spin qubit coherence time in hexagonal boron nitride. npj 2D Mater. Appl. 6, 60 (2022).

\bibitem{ref40} M. Onizhuk, K. C. Miao, J. P. Blanton, H. Ma, C. P. Anderson, A. Bourassa, D. D. Awschalom, and G. Galli, Probing the coherence of solid-state qubits at avoided crossings. PRX Quantum 2, 010311 (2021).

\bibitem{ref41} S. L. Bayliss, P. Deb, D. W. Laorenza, M. Onizhuk, G. Galli, D. E. Freedman, and D. D. Awschalom, Enhancing spin coherence in optically addressable molecular qubits through host-matrix control. Phys. Rev. X 12, 031028 (2022).

\bibitem{ref42} L. C. Bassett, A. Alkauskas, A. L. Exarhos, and K.-M. C. Fu, Quantum defects by design. Nanophotonics 8, 1867--1888 (2019).

\bibitem{ref43} G. Zhang, Y. Cheng, J.-P. Chou, and A. Gali, Material platforms for defect qubits and single-photon emitters. Appl. Phys. Rev. 7, 031308 (2020).

\bibitem{ref44} M. Widmann, S.-Y. Lee, T. Rendler, N. T. Son, H. Fedder, S. Paik, L.-P. Yang, N. Zhao, S. Yang, I. Booker, A. Denisenko, M. Jamali, S. A. Momenzadeh, I. Gerhardt, T. Ohshima, A. Gali, E. Janzén, and J. Wrachtrup, Coherent control of single spins in silicon carbide at room temperature. Nat. Mater. 14, 164--168 (2015).

\bibitem{ref45} D. B. Higginbottom, A. T. K. Kurkjian, C. Chartrand, M. Kazemi, N. A. Brunelle, E. R. MacQuarrie, J. R. Klein, N. R. Lee-Hone, J. Stacho, M. Ruether, C. Bowness, L. Bergeron, A. DeAbreu, S. R. Harrigan, J. Kanaganayagam, D. W. Marsden, T. S. Richards, L. A. Stott, S. Roorda, K. J. Morse, M. L. W. Thewalt, and S. Simmons, Optical observation of single spins in silicon. Nature 607, 266--270 (2022).

\bibitem{ref46} N. Chejanovsky, A. Mukherjee, J. Geng, Y.-C. Chen, Y. Kim, A. Denisenko, A. Finkler, T. Taniguchi, K. Watanabe, D. B. R. Dasari, P. Auburger, A. Gali, J. H. Smet, and J. Wrachtrup, Single-spin resonance in a van der Waals embedded paramagnetic defect. Nat. Mater. 20, 1079--1084 (2021).

\bibitem{ref47} S. Kanai, F. J. Heremans, H. Seo, G. Wolfowicz, C. P. Anderson, S. E. Sullivan, G. Galli, D. D. Awschalom, and H. Ohno, Generalized scaling of spin qubit coherence in over 12,000 host materials. Proc. Natl. Acad. Sci. U.S.A. 119, e2121808119 (2022).

\bibitem{ref48} O. Neitzke, A. J. Morfa, J. Wolters, A. W. Schell, G. Kewes, and O. Benson, Investigation of line width narrowing and spectral jumps of single stable defect centers in ZnO at cryogenic temperature. Nano Lett. 15, 3024-3029 (2015).

\bibitem{ref49} N. R. Jungwirth, H.-S. Chang, M. Jiang, and G. D. Fuchs, Polarization spectroscopy of defect-based single photon sources in ZnO. ACS Nano 10, 1210-1215 (2016).

\bibitem{ref50} A. J. Morfa, B. C. Gibson, M. Karg, T. J. Karle, A. D. Greentree, P. Mulvaney, and S. Tomljenovic-Hanic, Single-photon emission and quantum characterization of zinc oxide defects. Nano Lett. 12, 949--954 (2012).

\bibitem{ref51} M. Winter,The periodic table of the elements. \url{https://www.webelements.com} (2021).

\bibitem{ref52} A. Kołodziejczak-Radzimska and T. Jesionowski, Zinc Oxide---From Synthesis to Application: A Review. Materials 7, 2833-2881 (2014).

\bibitem{ref53} X. Linpeng, M. L. K. Viitaniemi, A. Vishnuradhan, Y. Kozuka, C. Johnson, M. Kawasaki, and K.-M. C. Fu, Coherence properties of shallow donor qubits in ZnO. Phys. Rev. Appl. 10, 064061 (2018).

\bibitem{ref54} V. Niaouris, M. V. Durnev, X. Linpeng, M. L. K. Viitaniemi, C. Zimmermann, A. Vishnuradhan, Y. Kozuka, M. Kawasaki, and K.-M. C. Fu, Ensemble spin relaxation of shallow donor qubits in ZnO. Phys. Rev. B 105, 195202 (2022).

\bibitem{ref55} X. Wang, C. Zimmermann, M. Titze, V. Niaouris, E. R. Hansen, S. H. D\textquotesingle Ambrosia, L. Vines, E. S. Bielejec, and K.-M. C. Fu, Properties of donor qubits in ZnO formed by indium-ion implantation. Phys. Rev. Appl. 19, 054090 (2023).

\bibitem{ref56} S. Zhang, T. Park, E. Perez, K. Li, X. Wang, M. Mansouri, Y. Wang, J. D. Vega Bazantes, R. Zhang, J. Sun, K.-M. C. Fu, H. Seo, and Y. Ping, Deep Spin Defects in Zinc Oxide for High-Fidelity Single-Shot Readout. PRX Quantum \textbf{7}, 020348 (2026).

\bibitem{ref57} A. Bilgin, I. N. Hammock, J. Estes, P. Udvarhelyi, R. Banal, J. R. Weber, G. Galli, and A. A. High, Donor-acceptor pairs in wide-bandgap semiconductors for quantum technology applications. npj Comput. Mater. 10, 7 (2024).

\bibitem{ref58} A. Bilgin, I. N. Hammock, A. A. High, and G. Galli, Donor-Acceptor Pairs Near Silicon Carbide Surfaces. J. Phys. Chem. Lett. 16, 10371-10380 (2025).

\bibitem{ref59} E. R. Hansen, V. Niaouris, B. E. Matthews, C. Zimmermann, X. Wang, R. Kolodka, L. Vines, S. R. Spurgeon, and K.-M. C. Fu, Isolation of single donors in ZnO. Phys. Rev. Lett. 133, 146902 (2024).

\bibitem{ref60} V. Niaouris, S. H. D\textquotesingle Ambrosia, C. Zimmermann, X. Wang, E. R. Hansen, M. Titze, E. S. Bielejec, and K.-M. C. Fu, Contributions to the optical linewidth of shallow donor-bound excitonic transition in ZnO. Optica Quantum 2, 7-13 (2024).

\bibitem{ref61} S. B. Orlinskii, J. Schmidt, P. G. Baranov, V. Lorrmann, I. Riedel, D. Rauh, and V. Dyakonov, Identification of shallow Al donors in Al-doped ZnO nanocrystals: EPR and ENDOR spectroscopy. Phys. Rev. B 77, 115334 (2008).

\bibitem{ref62} P. Li and W. Luo, Properties of carbon impurities in ZnO: A hybrid functional study. Phys. Rev. B 94, 075202 (2016).

\bibitem{ref63} W. A. Harrison, Electronic Structure and the Properties of Solids: The Physics of the Chemical Bond. Dover Publications, New York, 1989.

\bibitem{ref64} C. Freysoldt, B. Grabowski, T. Hickel, J. Neugebauer, G. Kresse, A. Janotti, and C. G. Van de Walle, First-principles calculations for point defects in solids. Rev. Mod. Phys. 86, 253-305 (2014).

\bibitem{ref65} C. Freysoldt, J. Neugebauer, and C. G. Van de Walle, Fully ab initio finite-size corrections for charged-defect supercell calculations. Phys. Rev. Lett. 102, 016402 (2009).

\bibitem{ref66} R. Yang, F. Wang, J. Lu, Y. Lu, B. Lu, S. Li, and Z. Ye, ZnO with \emph{p}-type doping: Recent approaches and applications. ACS Appl. Electron. Mater. 5, 4014--4034 (2023).

\bibitem{ref67} A. Janotti and C. G. Van de Walle, Native point defects in ZnO. Phys. Rev. B 76, 165202 (2007).

\bibitem{ref68} F. Oba, M. Choi, A. Togo, and I. Tanaka, Point defects in ZnO: an approach from first principles. Sci. Technol. Adv. Mater. 12, 034302 (2011).

\bibitem{ref69} P. Deák, B. Aradi, M. Kaviani, T. Frauenheim, and Á. Gali, Formation of NV centers in diamond: A theoretical study based on calculated transitions and migration of nitrogen and vacancy related defects. Phys. Rev. B 89, 075203 (2014).

\bibitem{ref70} L. Gordon, A. Janotti, and C. G. Van de Walle, Defects as qubits in 3C- and 4H-SiC. Phys. Rev. B 92, 045208 (2015).

\bibitem{ref71} K. Yim, J. Lee, D. Lee, et al., Property database for single-element doping in ZnO obtained by automated first-principles calculations. Sci. Rep. 7, 40907 (2017).

\bibitem{ref72} B. Kaduk, T. Kowalczyk, and T. Van Voorhis, Constrained Density Functional Theory. Chem. Rev. 112, 321-370 (2012).

\bibitem{ref73} F. Wu, D. Rocca, and Y. Ping, Dimensionality and anisotropicity dependence of radiative recombination in nanostructured phosphorene. J. Mater. Chem. C 7, 12891-12899 (2019).

\bibitem{ref74} T. J. Smart, K. Li, J. Xu, and Y. Ping, Intersystem crossing and exciton-defect coupling of spin defects in hexagonal boron nitride. npj Comput. Mater. 7, 59 (2021).

\bibitem{ref75} Y. Ma, M. Rohlfing, and A. Gali, Excited states of the negatively charged nitrogen-vacancy color center in diamond. Phys. Rev. B 81, 041204(R) (2010).

\bibitem{ref76} V. Ivády, G. Barcza, G. Thiering, S. Li, H. Hamdi, J.-P. Chou, Ö. Legeza, and Á. Gali, Ab initio theory of the negatively charged boron vacancy qubit in hexagonal boron nitride. npj Comput. Mater. 6, 41 (2020).

\bibitem{ref77} K. Li, V. D. Dergachev, I. D. Dergachev, S. Zhang, S. A. Varganov, and Y. Ping, Excited-state dynamics and optically detected magnetic resonance of solid-state spin defects from first principles. Phys. Rev. B 110, 184302 (2024).

\bibitem{ref78} J. R. Reimers, J. Shen, M. Kianinia, C. Bradac, I. Aharonovich, M. J. Ford, and P. Piecuch, Photoluminescence, photophysics, and photochemistry of the V\textsubscript{B}\textsuperscript{-} defect in hexagonal boron nitride. Phys. Rev. B 102, 144105 (2020).

\bibitem{ref79} Y. Jin, V. W.-Z. Yu, M. Govoni, A. C. Xu, and G. Galli, Excited state properties of point defects in semiconductors and insulators investigated with time-dependent density functional theory. J. Chem. Theory Comput. 19, 8689-8705 (2023).

\bibitem{ref80} A. Gali, E. Janzén, P. Deák, G. Kresse, and E. Kaxiras, Theory of spin-conserving excitation of the N-V- center in diamond. Phys. Rev. Lett. 103, 186404 (2009).

\bibitem{ref81} A. V. Ivanov, Y. L. A. Schmerwitz, G. Levi, and H. Jónsson, Electronic excitations of the charged nitrogen-vacancy center in diamond obtained using time-independent variational density functional calculations. SciPost Phys. 15, 009 (2023).

\bibitem{ref82} K. Huang and A. Rhys, Theory of light absorption and non-radiative transitions in F-centres. Proc. R. Soc. Lond. A 204, 406-423 (1950).

\bibitem{ref83} J. J. Markham, Interaction of normal modes with electron traps. Rev. Mod. Phys. 31, 956-989 (1959).

\bibitem{ref84} A. Alkauskas, J. L. Lyons, D. Steiauf, and C. G. Van de Walle, First-principles calculations of luminescence spectrum line shapes for defects in semiconductors: The example of GaN and ZnO. Phys. Rev. Lett. 109, 267401 (2012).

\bibitem{ref85} A. Alkauskas, B. B. Buckley, D. D. Awschalom, and C. G. Van de Walle, First-principles theory of the luminescence lineshape for the triplet transition in diamond NV centres. New J. Phys. 16, 073026 (2014).

\bibitem{ref86} F. Wu, T. J. Smart, J. Xu, and Y. Ping, Carrier recombination mechanism at defects in wide band gap two-dimensional materials from first principles. Phys. Rev. B 100, 081407(R) (2019).

\bibitem{ref87} A. Alkauskas, Q. Yan, and C. G. Van de Walle, First-principles theory of nonradiative carrier capture via multiphonon emission. Phys. Rev. B 90, 075202 (2014).

\bibitem{ref88} M. E. Turiansky, K. Parto, G. Moody, and C. G. Van de Walle, Rational design of efficient defect-based quantum emitters. APL Photonics 9, 066117 (2024).

\bibitem{ref89} G. Kresse and J. Furthmüller, Efficiency of ab-initio total energy calculations for metals and semiconductors using a plane-wave basis set. Comput. Mater. Sci. 6, 15-50 (1996).

\bibitem{ref90} J. Heyd, G. E. Scuseria, and M. Ernzerhof, Hybrid functionals based on a screened Coulomb potential. J. Chem. Phys. 118, 8207-8215 (2003).

\bibitem{ref91} P. E. Blöchl, Projector augmented-wave method. Phys. Rev. B 50, 17953-17979 (1994).

\bibitem{ref92} V. Wang, N. Xu, J.-C. Liu, G. Tang, and W.-T. Geng, VASPKIT: A user-friendly interface facilitating high-throughput computing and analysis using VASP code. Comput. Phys. Commun. 267, 108033 (2021).

\bibitem{ref93} N. Ashkenov, B. N. Mbenkum, C. Bundesmann, V. Riede, M. Lorenz, D. Spemann, E. M. Kaidashev, A. Kasic, M. Schubert, M. Grundmann, and G. Wagner, Infrared dielectric functions and phonon modes of high-quality ZnO films. J. Appl. Phys. 93, 126-133 (2003).

\bibitem{ref94} A. Gali, M. Fyta, and E. Kaxiras, Ab-initio supercell calculations on nitrogen-vacancy center in diamond: Electronic structure and hyperfine tensors. Phys. Rev. B 77, 155206 (2008).

\bibitem{ref95} T. A. Abtew, Y. Y. Sun, B.-C. Shih, P. Dev, S. B. Zhang, and P. Zhang, Dynamic Jahn-Teller effect in the NV- center in diamond. Phys. Rev. Lett. 107, 146403 (2011).

\bibitem{ref96} S. L. Dudarev, G. A. Botton, S. Y. Savrasov, C. J. Humphreys, and A. P. Sutton, Electron-energy-loss spectra and the structural stability of nickel oxide: An LSDA+U study. Phys. Rev. B 57, 1505-1509 (1998).

\bibitem{ref97} A. Marini, C. Hogan, M. Grüning, and D. Varsano, yambo: An ab initio tool for excited state calculations. Comput. Phys. Commun. 180, 1392-1403 (2009).

\bibitem{ref98} D. Sangalli, A. Ferretti, H. Miranda, C. Attaccalite, I. Marri, E. Cannuccia, P. Melo, M. Marsili, F. Paleari, A. Marrazzo, G. Prandini, P. Bonfà, M. O. Atambo, F. Affinito, M. Palummo, A. Molina-Sánchez, C. Hogan, M. Grüning, D. Varsano, and A. Marini, Many-body perturbation theory calculations using the yambo code. J. Phys.: Condens. Matter 31, 325902 (2019).

\bibitem{ref99} P. Giannozzi, S. Baroni, N. Bonini, M. Calandra, R. Car, C. Cavazzoni, D. Ceresoli, G. L. Chiarotti, M. Cococcioni, I. Dabo, A. Dal Corso, S. de Gironcoli, S. Fabris, G. Fratesi, R. Gebauer, U. Gerstmann, C. Gougoussis, A. Kokalj, M. Lazzeri, L. Martin-Samos, N. Marzari, F. Mauri, R. Mazzarello, S. Paolini, A. Pasquarello, L. Paulatto, C. Sbraccia, S. Scandolo, G. Sclauzero, A. P. Seitsonen, A. Smogunov, P. Umari, and R. M. Wentzcovitch, QUANTUM ESPRESSO: a modular and open-source software project for quantum simulations of materials. J. Phys.: Condens. Matter 21, 395502 (2009).

\bibitem{ref100} D. R. Hamann, Optimized norm-conserving Vanderbilt pseudopotentials. Phys. Rev. B 88, 085117 (2013).

\bibitem{ref101} M. Schlipf and F. Gygi, Optimization algorithm for the generation of ONCV pseudopotentials. Comput. Phys. Commun. 196, 36-44 (2015).

\bibitem{ref102} M. Govoni and G. Galli, Large scale GW calculations. J. Chem. Theory Comput. 11, 2680-2696 (2015).

\bibitem{ref103} Y. Jin, M. Govoni, and G. Galli, Vibrationally resolved optical excitations of the nitrogen-vacancy center in diamond. npj Comput. Mater. 8, 238 (2022).

\bibitem{ref104} M. E. Turiansky, A. Alkauskas, M. Engel, G. Kresse, D. Wickramaratne, J.-X. Shen, C. E. Dreyer, and C. G. Van de Walle, Nonrad: Computing nonradiative capture coefficients from first principles. Comput. Phys. Commun. 267, 108056 (2021).

\end{thebibliography}
\end{document}